\documentclass[prl,aps,amssymb,amsfonts,amsmath,showpacs,twocolumn,superscriptaddress,tightenlines]{revtex4}

\usepackage{graphicx}

\begin{document}
\newcommand{\ethaffil}{Laboratory of Physical Chemistry, Swiss Federal Institute of Technology (ETH), 8093 Zurich, Switzerland}
\newcommand{\mariaaffil}{Institute of Electronic Structure and Laser (IESL), Foundation for
Research and Technology Hellas (FORTH), P.O. Box 1527, 71110
Heraklion, Crete, Greece}

\newcommand{\figwidth}{0.45\textwidth}
\newcommand{\real}{\operatorname{Re}}

\title{Tuning of High-Q Photonic Crystal Microcavities by a Subwavelength Near-field Probe}

\author{A. Femius {Koenderink}}
\affiliation{\ethaffil}
\author{Maria Kafesaki}
\affiliation{\mariaaffil}
\author{Ben  C. Buchler}
\affiliation{\ethaffil}
\author{Vahid Sandoghdar}\email{vahid.sandoghdar@ethz.ch}
\affiliation{\ethaffil}

\date{Prepared for Phys. Rev. December 15, 2004}

\begin{abstract}
We demonstrate that the resonance frequencies of high-Q
microcavities in two-dimensional photonic crystal membranes can be
tuned over a wide range by introducing a subwavelength dielectric
tip into the cavity mode. Three-dimensional finite-difference
time-domain (FDTD) simulations show that by varying the lateral
and vertical positions of the tip, it is possible to tune the
resonator frequency without lowering the quality factor. Excellent
agreement with a perturbative theory is obtained, showing that the
tuning range is limited by the ratio of the cavity mode volume to
the effective polarizability of the nano-perturber.
\end{abstract}

\pacs{42.70.Qs, 42.50.Pq, 42.60.Da, 42.64.Pq, 42.25.Fx.}

\maketitle

Solid-state optical microresonators are of great interest to a
wide range of fields such as bio-sensing, nonlinear optics,
low-threshold lasers and cavity quantum
electro-dynamics~\cite{microcavity,vahala}. Two cavity properties
that are commonly desired,  but often incompatible,  are very high
quality factors (Q) and very small mode volumes. Photonic crystal
microcavities in  thin semiconductor
membranes~\cite{vuckovic,nodacav,srinivasan} promise a good
compromise between these features. One of the difficulties
associated with such monolithic ultrasmall resonators is to match
their resonance frequencies with those of interest in a given
application. This is especially a concern since fabrication
tolerances make it nearly impossible to realize the exact design
parameters.  Temperature tuning can sometimes be used  to meet a
resonance condition. For example, Yoshie et al. used this
technique to control the coupling of a quantum dot with a
microcavity~\cite{yoshie}.  This was possible due to the
temperature sensitivity of the quantum dot. In general, however,
it is desirable to independently tune  the cavity resonance
without manipulating the system that it couples to.  In this
Letter we show that the introduction of a subwavelength dielectric
object, such as a scanning probe, can achieve this.

\begin{figure*}
\includegraphics[width=\textwidth]{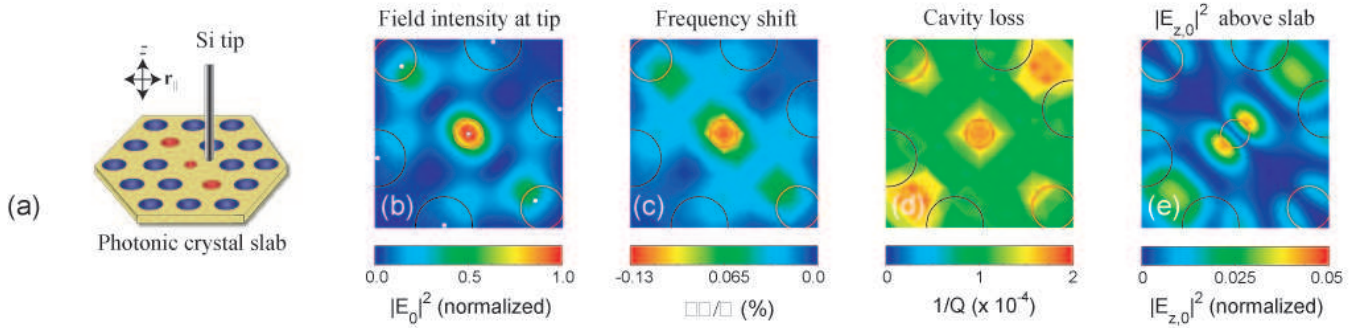}
\caption{(Color) (a) Schematic of the system under study. A
cylindrical silicon tip with a diameter of 125 nm is placed at
height $z_{\rm tip}$ and lateral position $\mathbf{r}_{||}$ in
front of a photonic crystal slab. Reduction of diameters for 3
pores (shaded red) and shift of two of them yields a cavity with a
resonance at $\lambda=1500$~nm and a quality factor of 13000. (b)
Intensity distribution $|\mathbf{E}_0|^2$ of the unperturbed mode
in a plane $30$~nm above the slab. The black circles display the
positions of pores (red circles for tuned pores) whereas the small
white dots mark the original locations of the pores in the
hexagonal lattice. (c) Calculated resonance frequency shift
induced by the tip placed 30~nm above the slab, as a function of
its lateral position. (d) The inverse $Q$ corresponding to (c).
The unperturbed cavity has $1/Q=7.5\cdot 10^{-5}$. (e)
Contribution of the $z$-component $|E_{z,0}|^2$ to the unperturbed
field intensity, integrated over the extent of the tip above the
slab. Field intensities in (b) and (e) are normalized to the
maximum total field intensity $|\mathbf{E}_{0}|^2$.
 \label{fig:cartoon}\label{fig:shift}}
\end{figure*}

A Scanning Near-field Optical Microscope (SNOM) tip has recently
been used to image the intensity distribution in a low-Q photonic
crystal microresonator~\cite{kramper}. The central idea in our
current work is to investigate how the presence of an external
dielectric nanometer-sized object, such as a SNOM tip, can modify
the resonance condition of a photonic crystal microresonator (see
Fig.~\ref{fig:cartoon}(a)). This scheme is analogous to a
technique commonly used in microwave engineering whereby the
frequency of a resonator is adjusted by insertion of dielectric
stubs ~\cite{waldron}. In that case,  because microwave resonators
are typically closed, scattering from the object does not cause
any loss, and the cavity Q is determined by absorption. However,
when a dielectric object is placed into the field of an open
cavity, such as a photonic crystal slab resonator, scattering
dominates. Indeed, it has been shown experimentally that coupling
of a glass fiber tip can shift the resonance frequency of a high-Q
microresonator at the cost of reducing its quality
factor~\cite{goetzinger-APB}. Since a photonic crystal microcavity
has a very small mode volume and relies on the precise arrangement
of dielectric material at the subwavelength scale, one might
expect that introducing the slightest external object would spoil
the quality factor. In what follows we show that this is not
necessarily the case. It turns out that a small dielectric object
acts in the same fashion as an atom which, within classical linear
dispersion theory, changes the frequency of a high-finesse cavity
according to its polarizability and its position relative to the
nodes of the cavity mode~\cite{berman,zhu90}.

We consider a dielectric cylindrical tip near a membrane-type
photonic crystal, consisting of a high index slab that is
perforated with a hexagonal lattice (lattice constant $a=420$~nm)
and surrounded by air.  We assume a slab dielectric constant
$\epsilon=11.76$, a thickness of 250~nm, and a hole radius
$r=0.3a$.  As  Fig.~\ref{fig:cartoon}(a) shows, the resonator is
formed about a defect of reduced radius $r=0.15a$, similar to the
optimization in Ref.~\cite{vuckovic}. By reducing the radius of
just two holes to $r=0.23a$ on either side of the defect and then
shifting them outwards by $0.11a$, we create a nondegenerate
dipole mode with a mode profile as shown in
Fig.~\ref{fig:shift}(b). The mode has a frequency $\omega a/2\pi
c=0.284$ in the center of the 2D band gap.  The $Q$ is around
$13000$, corresponding to a resonance linewidth of 15~GHz at
$\lambda= 1500$~nm. Crystals over $5\times 5~\mu$m in lateral size
surrounded by up to $1~\mu$m of air were simulated using the 3D
FDTD method with Liao's absorbing boundary
conditions~\cite{fdtd,fan}. Computational meshes were 14 or 20
grid points per $a$ parallel to the membrane and had doubled
resolution normal to the membrane~\cite{kafesaki}. Grid-cell
volume averaging of the dielectric constant was employed to reduce
staircasing errors~\cite{hermann}. Quality factors and cavity mode
frequencies were obtained by fitting a damped harmonic wave to
time traces of the total $E$-field energy in the cavity.

\begin{figure}
\centerline{\includegraphics[width=\figwidth]{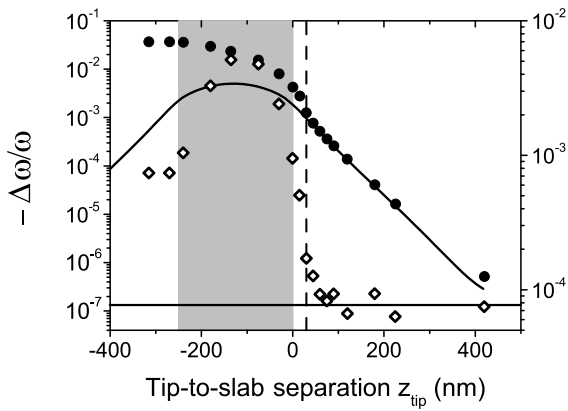}}
\caption{Relative frequency shift (solid symbols, left hand axis)
and inverse $Q$ (open symbols, right hand axis) as a function of
the separation $z_{\rm tip}$ between the surface of the photonic
crystal and a silicon tip of 125~nm diameter aligned with the
central defect. The grey area indicates the extent of the
membrane. For $z_{\rm tip}<0$ the tip extends into or through the
defect. The dashed line indicates $z_{\rm tip}=30$~nm
corresponding to the data in Figs.~\ref{fig:shift}(b--d). The
black curve (scaled) shows the unperturbed mode profile
$|\mathbf{E}_0|^2$. The horizontal  line indicates $1/Q$ in
absence of the tip. \label{fig:centershift}}
\end{figure}

Figure~\ref{fig:shift}(c) displays a contour plot of the relative
frequency shift $\Delta\omega/\omega$ as a function of the lateral
position of a silicon tip of 125 nm diameter,  placed at a height
of $30$~nm above the photonic crystal slab. As expected, the
introduction of high index material in the cavity mode causes a
red shift of the mode frequency, depending on the tip position.
The maximum relative frequency shift of $0.13$\% amounts to about
2~nm (or 260~GHz) at $\lambda=1500$~nm and occurs when the tip is
placed above  the central defect. At a lateral distance of 200~nm
from the central hole, the shift is already less than 10\% of this
maximum value, but as the tip approaches the two shifted holes,
again a large tuning of up to 200~GHz can be achieved.
Figure~\ref{fig:centershift} displays the dependence of the cavity
resonance on the separation between the tip apex and the photonic
crystal slab ($z_{\rm tip}$), for the case where the tip is
aligned with the cavity center. The filled circles show that as
the tip is approached from afar ($z_{\rm tip}>0$), the frequency
shift grows exponentially with a $1/e$ length of $d=50$~nm.  When
the tip is put through the slab ($z_{\rm tip}<0$), the increasing
amount of dielectric material inserted in the mode profile
continuously detunes the resonance to the red, saturating at a
shift of $\Delta\omega/\omega=-3.7\%$. The black curve in
Fig.~\ref{fig:centershift} displays the variation of the
unperturbed cavity mode intensity  $|\mathbf{E}_0|^2$. The
excellent agreement between this curve and the filled symbols for
($z_{\rm tip}>0$) indicates that the frequency shift maps
$|\mathbf{E}_0|^2$. This direct correspondence of the frequency
shift to the mode profile is also evident from the strong
similarity of the lateral images in Figs.~\ref{fig:shift}(b) and
(c).

An exciting aspect of cavity tuning with a tip is that a large
frequency shift can be achieved without inducing a considerable
cavity loss. Open symbols in Fig.~\ref{fig:centershift} show that
within the accuracy of the simulations the cavity $Q$ of 13000 has
not been affected at separations down to about 60~nm while the
cavity resonance has been tuned by 100 GHz. Even at a height of
30~nm where  $\Delta \omega$=260~GHz,  the
cavity $Q$ remains as high as 5000. If the tip is pushed closer to
or into the central defect hole, however, the continuous increase
of the frequency shift is accompanied by a strong reduction in
$Q$. In this regime, the tip perturbs the mode profile and strong
out-of-plane scattering occurs.  When the tip extends completely
through the defect slab and the distribution of dielectric
material becomes more symmetric, the $Q$ recovers to about $1000$.

To gain quantitative insight into the modification of the cavity
resonance,   we have taken advantage of the established theory of
dielectric perturbations in microwave
cavities~\cite{waldron,inoue}.  Suppose the tip has a dielectric
constant of  $\epsilon_{p}$ and is contained within a volume
$V_{p}$.  Maxwell's relations imply the \emph{exact} expression,
\begin{equation}
\frac{\Delta {\omega}}{\omega}-\frac{i }{2 \Delta Q}=
\displaystyle \frac{-\int_{ V_{p}}
[\epsilon_{p}-\epsilon(\mathbf{r})]
\real[\mathbf{E}^*_0\cdot\mathbf{E}_p] d\mathbf{r}+\frac{
i}{\omega} \oint_{\delta V} \real[\mathbf{S}_p] \cdot \mathbf{n}
da }{\int_{V} \real[\mathbf{E}^*_0\cdot\mathbf{D}_p
+\mathbf{H}^*_0\cdot\mathbf{B}_p] d\mathbf{r}}
\label{eq:exact}\end{equation} for the relative frequency shift
($\Delta\omega/\omega$) and induced scattering loss  ($\Delta
Q^{-1}$).  Here $\epsilon(\mathbf{r})$ is the dielectric constant
of the unperturbed system and $\mathbf{E}_{0(p)}$, and
$\mathbf{H}_{0(p)}$ are the complex electric and magnetic fields
of the unperturbed (perturbed) system.  The integral in the
denominator runs over some volume $V$ enclosing the entire system.
The induced scattering loss is proportional to the integrated flux
of $\mathbf{S}_p=\mathbf{E}^*_p\times \mathbf{H}_0
-\mathbf{E}^*_0\times \mathbf{H}_p$ through the outer surface
$\delta V$ of $V$.  We note that this dependence is very different
to that of the frequency shift that is proportional to an integral
over the perturbing volume $V_{p}$.

\begin{figure}
\centerline{\includegraphics[width=\figwidth]{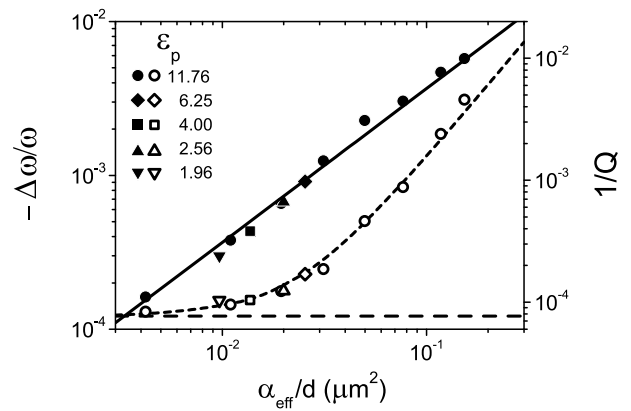}}
\caption{Relative frequency shift (solid symbols, left hand axis)
and inverse $Q$ (open symbols, right hand axis) as a function of
the effective polarizability  per length $\alpha_{\rm
eff}/d=3(\epsilon_{\rm  p}-1)/(\epsilon_{\rm  p}+2) \pi r_{\rm
 p}^2$, for  cylindrical tips positioned 30~nm above the central
defect in Fig.~1. Symbol shapes indicate various tip materials
(Si, TiO$_2$, ZrO$_2$, polystyrene, and SiO$_2$ in order of
decreasing $\epsilon_{\rm  p}$ as indicated). The frequency shift
is proportional to $\alpha_{\rm eff}$ (solid curve). The loss
$1/Q$ can be fitted (dashed curve) with the sum of the loss
without the tip (horizontal dashed line) and a contribution
proportional to $\alpha_{\rm eff}^2$. \label{fig:polariz}}
\end{figure}

A simple picture for the frequency shift emerges in the weakly
perturbative regime  where $\frac{\Delta {\omega}}{\omega} \ll 1$
and the $Q$ is not degraded, in our case corresponding to the
range $z_{\rm tip}\geq 30$~nm. In this regime, the denominator in
Eq.~(\ref{eq:exact}) is independent of the perturbation and the
frequency shift is set by the overlap integral of the perturbed
and unperturbed fields within the tip volume $V_p$ only. The field
$\mathbf{E}_p$ in the tip is given by
$\mathbf{E}_p=3\mathbf{E}_0/(\epsilon_{p} +2)$, where the
proportionality constant takes into account the local field
effects~\cite{inoue}. Next, we separate the exponential
$z$-dependence of
$\mathbf{E}_0=\mathbf{E}_0(\mathbf{r}_{||})e^{-z/2d}$  from its
dependence on the lateral tip coordinate $\mathbf{r}_{||}$, and
assume that the unperturbed field $\mathbf{E}_0(\mathbf{r}_{||})$
is constant over the small tip cross section. We then find
\begin{equation}
\frac{\Delta \omega (\mathbf{r}_{||},z_{\rm tip})}{\omega}= -
\frac{\alpha_{\rm eff} }{2V_{\rm
cav}}\frac{|E_0(\mathbf{r}_{||})|^2}{\max
[\epsilon(\mathbf{r})|E_0|^2]}e^{-z_{\rm tip}/{d }}.
\label{eq:approx}\end{equation} Here $V_{\rm cav}=\int
\epsilon(\mathbf{r})|E_0|^2 d\mathbf{r}/\max[\epsilon(\mathbf{r})
|E_0|^2]$ is defined as the cavity mode
volume~\cite{inoue,vuckovic} and $\alpha_{\rm eff}$ is the
effective polarizability of the tip
\begin{equation}
\alpha_{\rm eff}= 3\frac{\epsilon_{p}-1}{\epsilon_{p}+2} V_{\rm
eff}, \label{eq:electrostat}
\end{equation} equal to the electrostatic
polarizability of  a volume $V_{\rm eff}$ of material with
dielectric constant $\epsilon_{p}$~\cite{jackson}. The exponential
decay of the cavity mode (with decay length $d$) limits the volume
of the tip that contributes to the polarizability to $V_{\rm eff}
=\pi r_{p}^2 d$ for a  tip of radius $r_p$~\cite{lengthref}. This
analysis reproduces all the features of our FDTD results
concerning the frequency shift. Firstly, it yields the exponential
decrease of $\Delta\omega/\omega$ as a function of the tip-slab
separation $z_{\rm tip}$ (see Fig.~\ref{fig:centershift}).
Secondly, it confirms the direct correspondence between the
frequency shift and the unperturbed mode profile (see
Figs.~\ref{fig:shift}(b,c)). Thirdly,  Eq.~(\ref{eq:approx})
states that the induced frequency shift is inversely proportional
to the mode volume, a feature that we have confirmed by
simulations of different cavity designs (e.g. that of
Ref~\cite{nodacav}). Further confirmation is the quantitative
agreement of the frequency shifts in Fig.~\ref{fig:centershift}
with the mode volume $V_{\rm cav}=0.02~\mu$m$^3$ determined from
the unperturbed mode profile of our cavity. Finally, and most
significantly, Eq.~(\ref{eq:approx}) predicts that the frequency
shift is proportional to $\alpha_{\rm eff}$. To investigate this
we performed FDTD simulations for tips of various radii and
materials, fixed above the cavity center at $30$~nm. The closed
symbols and the solid line in Fig.~\ref{fig:polariz} demonstrate
that the frequency shift is proportional to the effective
polarizability of the tip. In other words, in the region
 $z_{\rm tip}>30$~nm the tip acts as a polarizable object that
is weakly coupled to the cavity field, much in the same manner as
an atom that can be described by a classical dipole of the
appropriate polarizability within   linear dispersion
theory~\cite{zhu90,berman}. This is a remarkable finding because
the tip is not much smaller than the central hole of the
microcavity and its mode volume, so that the validity of dipole
approximation is not a trivial matter.

Next we turn our attention to the origin of losses induced by the
tip. The power radiated by an object of polarizability
$\alpha_{\rm eff}$ can be modelled as a Rayleigh scattering
process~\cite{jackson}, resulting in a quadratic dependence on
$\alpha_{\rm eff}$. Indeed the open symbols in
Fig.~\ref{fig:polariz} display exactly this dependence when the
tip is at the center of the cavity. The quadratic dependence of
the loss compared to the linear dependence of the resonance
frequency on $\alpha_{\rm eff}$   makes it possible to tune the
frequency of high-$Q$ microcavities without incurring prohibitive
losses. Intriguingly, however, the induced loss is not directly
related to the strength of the unperturbed cavity mode at the
position of the tip.  A comparison of Figs.~\ref{fig:shift}(b) and
(d) shows that there are tip positions where the losses are high
even though the electric field intensity is small (upper right and
lower left corners).  It turns out, as shown in
Fig.~\ref{fig:shift}(e), that the cavity field at these points has
a large z-component ($E_{z,0}$). The magnitude of   $|E_{z,0}|^2$
at these points amounts to about 5\% of the maximum value of
$|\mathbf{E}_{0}|^2$ attained above the center of the cavity.
Substantial loss is induced  by   scattering of the field into the
tip, which is known to be much more efficient for the z-component
of the field
 than for the  field components perpendicular to the
tip~\cite{Girard-94}. FDTD simulations of the perturbed fields
further confirm the polarization-specific enhanced coupling into
the tip~\cite{supplementarymat}. The disparity between induced
loss and the cavity mode profile underlines a fundamental feature
of Eq.~(\ref{eq:exact}): contrary to the case of the frequency
shift, the scattering loss does not depend solely on the fields
within the perturbing volume only. Instead, it is the far-field
interference of the fields encoded in the integrated flux
$\mathbf{S}_p$ that is the decisive factor. In this sense, there
is a close relation between the perturbation of a photonic crystal
microcavity by a tip and the tuning of holes around the cavity as
used recently for the optimisation of the quality
factor~\cite{vuckovic,srinivasan,nodacav}.

In conclusion, we have shown that sharp dielectric tips in the
near-field can be used to tune photonic crystal microcavities over
a large range without inducing prohibitive losses.  Our analysis
shows that the tuning range scales with the ratio of the tip
polarizability to the mode volume. Hence, this scheme is expected
to be effective for any electromagnetic resonance localized in a
volume comparable to the perturbing polarizability, as is the case
in microspheres, microdisks or micropillars
~\cite{vahala,goetzinger-APB,yoshie}.
 One
advantage of this scheme is that the tuning process does not
influence the optical properties of the emitters embedded in the
cavity. Furthermore, the rapid progress in nanotechnology makes it
feasible to integrate a tip with a microcavity in a
device~\cite{cantilever}. Compared to existing methods for tuning
photonic crystals~\cite{yoshino},   tip tuning can create and
reverse the frequency shift more quickly and in contrast to
ultrafast optical tuning~\cite{leonard}, the frequency shift  can
be maintained   indefinitely. In other words, tip tuning does not
depend on the material properties of the crystal.
 The time scale for   tuning is only
limited by mechanical resonances of the tip which can reach the
MHz regime~\cite{cantilever}. This opens the possibility of using
a photonic crystal microcavity as an optical switch for integrated
optics applications.

This work was funded by the Deutsche Forschungsgemeinschaft (DFG)
through focus program SP1113 and by ETH Zurich. M.K. thanks Prof.
C. M. Soukoulis for continuous support and fruitful discussions.

\newpage

\begin{figure*}

\noindent EPAPS material corresponding to `\emph{Tuning of High-Q
Photonic Crystal Microcavities by a Sub-wavelength Near-field
Probe}' by A.F. Koenderink, M. Kafesaki, B.C. Buchler, and V.
Sandoghdar.

\vfill
\centerline{\includegraphics[width=0.85\textwidth]{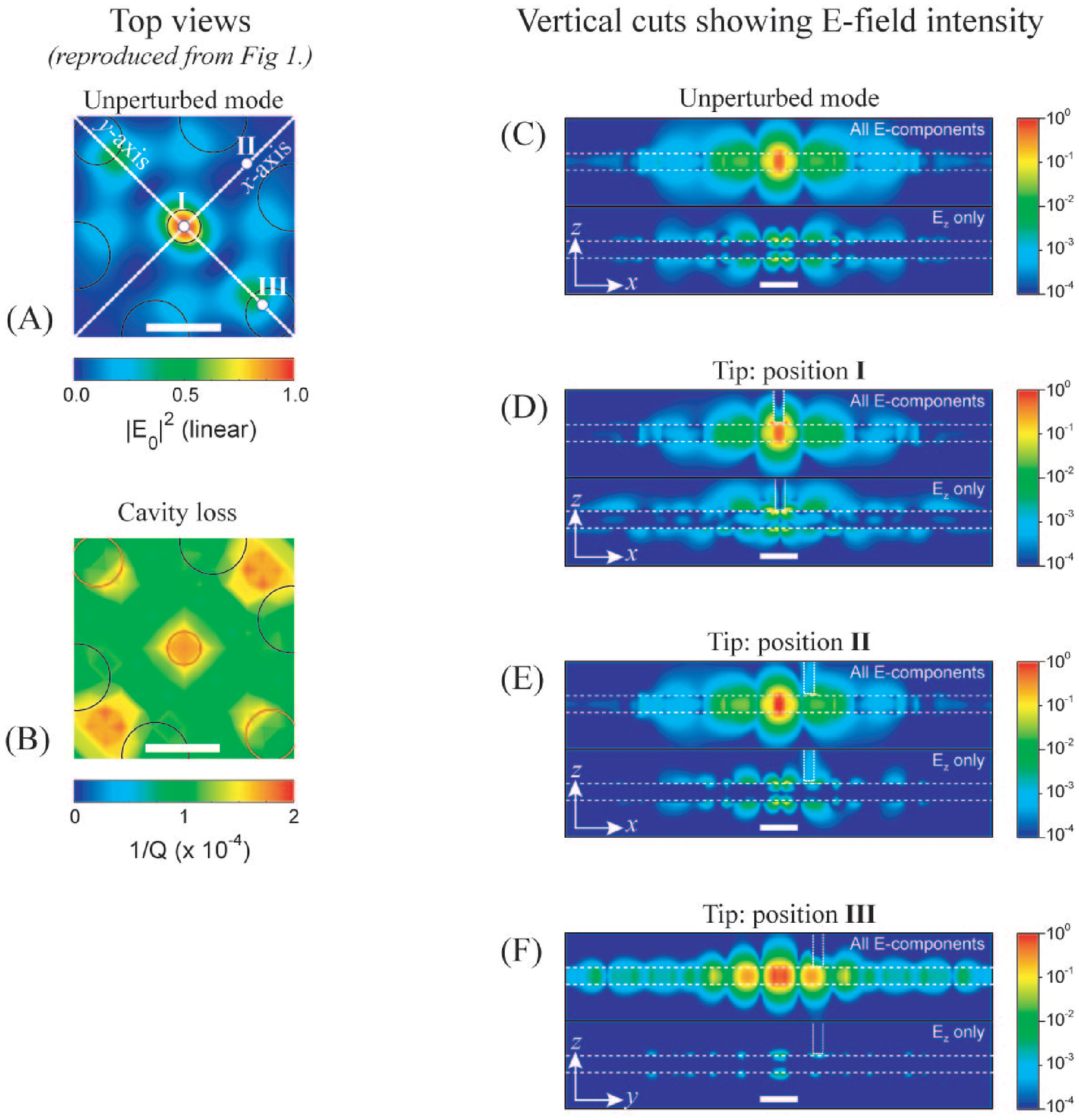}}

\vfill

\noindent (Color) Panel (A):  total E-field intensity of the
cavity mode in a plane 30~nm above the slab (reproduced from
manuscript, Fig. 1 (b)). Lines indicate the directions of the
crosscuts  along $x$ in (C)--(E) and along $y$ in (F), and circles
mark the tip positions in (D)--(F) as labelled.  Panel (B): cavity
loss versus tip position (manuscript, Fig. 1(d)). Panels (C)-(F):
Contour plots of the total E-field intensity
$|\mathbf{E}_0|^2/\max(|\mathbf{E}_0|^2)$ (top half of each panel)
and the contribution of the $z$-component E-field intensity $|
{E}_{z,0}|^2/\max(|\mathbf{E}_0|^2)$  (bottom half of each panel).
Color scales are logarithmic (shown on the right of the panels).
Field intensities are  plotted for vertical plane cuts through the
photonic crystal membrane along directions indicated by dotted
lines in (A). Compared to the unperturbed profile (C),
 panel (D) shows that if the tip is above the central hole (position
I in panel (A)) light does not couple into the tip, but scatters
above and below the slab. Panel (E) shows that efficient coupling
into the tip occurs when the tip is located in between two untuned
holes (position II in panel (A)), leading to a large reduction in
Q despite weak overall field intensity at position II. This
coupling into the tip is selective for lobes of the z-component of
the E-field, and so does not occur for position I amd III (panel
F). In all cases, Si tips of 125~nm diameter (dotted outlines in
(D)--(F)) were considered, at a height 30 nm above the crystal
slab (top and bottom face indicated by dashes). White bars in all
panels indicate $a=420$~nm.

\end{figure*}

\end{document}